\documentclass[12pt]{iopart}

\usepackage{graphics}

\begin{document}

\title[]{Unusual behaviours and Impurity Effects in the Noncentrosymmetric Superconductor CePt$_{3}$Si}

\author{I Bonalde$^1$, R L Ribeiro$^1$, W Br\"{a}mer-Escamilla$^1$,
C Rojas$^2$, E Bauer$^3$, A Prokofiev$^3$, Y Haga$^4$, T
Yasuda$^5$ and Y \={O}nuki$^5$}

\address{$^1$ Centro de F\'{\i}sica, Instituto Venezolano de
Investigaciones Cient\'{\i}ficas, Apartado 20632, Caracas
1020-A, Venezuela}

\address{$^2$ Departamento de F\'{\i}sica, Facultad de Ciencias,
Universidad Central de Venezuela, Apartado 47586, Caracas
1041-A, Venezuela}

\address{$^3$ Institut f\"{u}r Festk\"{o}rperphysik, Technische
Universit\"{a}t Wien, A-1040 Wien, Austria}

\address{$^4$ Advanced Science Research Center, Japan Atomic Energy
Research Institute, Tokai, Ibaragi 319-1195, Japan}

\address{$^5$ Graduate School of Science, Osaka University, Toyonaka,
Osaka 560-0043, Japan}

\ead{bonalde@ivic.ve}

\begin{abstract}
We report a study in which the effect of defects/impurities,
growth process, off-stoichiometry, and presence of impurity
phases on the superconducting properties of noncentrosymmetric
CePt$_3$Si is analysed by means of the temperature dependence
of the magnetic penetration depth. We found that the linear
low-temperature response of the penetration depth -indicative
of line nodes in this material- is robust regarding sample
quality, in contrast to what is observed in unconventional
centrosymmetric superconductors with line nodes. We discuss
evidence that the broadness of the superconducting transition
may be intrinsic, though not implying the existence of a second
superconducting transition. The superconducting transition
temperature systematically occurs around 0.75 K in our
measurements, in agreement with resistivity and ac magnetic
susceptibility data but in conflict with specific heat, thermal
conductivity and NMR data in which $T_c$ is about 0.5 K. Random
defects do not change the linear low-temperature dependence of
the penetration depth in the heavy-fermion CePt$_3$Si with line
nodes, as they do in unconventional centrosymmetric
superconductors with line nodes.
\end{abstract}

\maketitle

\section{Introduction}

In general, the superconducting BCS ground state is formed by
Cooper pairs with zero total angular momentum. The electronic
states are four-fold degenerate: $|\textbf{k}\uparrow\rangle$,
$|-\textbf{k}\uparrow\rangle$, $|\textbf{k}\downarrow\rangle$,
and $|-\textbf{k}\downarrow\rangle$ have the same energy
$\epsilon(\textbf{k})$. The states with opposite momenta and
opposite spins are transformed to one another under time
reversal operation $\hat{K}|\textbf{k}\uparrow\rangle =
|-\textbf{k}\downarrow\rangle$, and states with opposite
momenta are transformed to one another under inversion
operation $\hat{I}|\textbf{k}\downarrow\rangle =
|-\textbf{k}\downarrow\rangle$. The four degenerate states are
a consequence of space and time inversion symmetries. Parity
symmetry is irrelevant for spin-singlet pairing, but is
essential for spin-triplet pairing. Time reversal symmetry is
required for spin-singlet configuration, but is unimportant for
spin-triplet state \cite{anderson1,anderson2}.

All conventional superconductors ($s$-wave pairing states)  are
examples of systems invariant under parity and time reversal
symmetries. Some superconductors, like UPt$_3$ and
Sr$_2$RuO$_4$, violate time reversal symmetry and their Cooper
pairs form spin-triplet states. The heavy fermion CePt$_3$Si is
the classic example of superconductors without inversion
symmetry. Parity is not a good quantum number in
noncentrosymmetric superconductors, therefore it is not
possible to classify their pairing states as pure even-parity
spin-singlet or pure odd-parity spin-triplet. Rather, parity
mixing is expected. Moreover, the lack of inversion symmetry
causes the appearance of an antisymmetric spin-orbit coupling
(ASOC) that lifts the spin degeneracy existing in
parity-conserving superconducting systems by originating two
bands with different spin structures. From this it follows that
superconductors without inversion symmetry should show
intriguing properties, as indeed is the case for CePt$_3$Si.

Among the striking behaviours of CePt$_3$Si ($T_c=0.75$ K)
\cite{bauer} one finds: (a) a small peak just below the
superconducting transition in the NMR $1/T_1T$ data
\cite{yogi}, characteristic of spin-singlet superconductors
with a nodeless energy gap; (b) absence of Pauli paramagnetic
limiting field \cite{bauer}, indicative of spin-triplet pairing
which is forbidden by the lack of inversion symmetry; and (c)
low-temperature power-law behaviours in the magnetic
penetration depth \cite{mine7} and the thermal conductivity
\cite{izawa}, expected for energy gaps with line nodes.
However, these conflicting findings have been satisfactorily
explained by a model based on the splitting of the spin
degenerate band produced by the ASOC
\cite{frigeri,hayashi1,hayashi2}.

The superconducting phase of CePt$_3$Si appears to be very
complex \cite{mine8,mine11} and several intriguing issues
remain to be elucidated: (a) the broad transition in the
electrical resistivity of a clean sample \cite{yasuda}, (b) the
broad transition and a weak second drop around 0.5 K in
penetration depth \cite{mine7}, (c) a second peak around 0.5 K
in specific heat \cite{scheidt,jskim} and ac magnetic
susceptibility \cite{nakatsuji,aoki2}, and (d) the occurrence
of the superconducting transition at a rather low temperature
near 0.5 K in thermal conductivity and some specific heat and
NMR data \cite{izawa,tateiwa,takeuchi2,yogi2}. The broad
transition and second peak have been attributed to the presence
of antiferromagnetic impurity phases in the samples
\cite{jskim} and to deviations from the 1:3:1 nominal
stoichiometry and/or structural defects
\cite{motoyama,motoyama2}. The second peak has also been
interpreted as the indication of a second superconducting
transition \cite{scheidt,nakatsuji}. The superconducting
transition has even been suggested to actually occur around 0.5
K instead of at 0.75 K \cite{takeuchi,motoyama2}.

The unusual/conflicting results discussed above seem somewhat
sample dependent. However, two of these results have been found
in different types of samples: The small peak below the
critical temperature in NMR $1/T_1T$ data \cite{yogi,yogi2} and
the absence of a paramagnetic limiting field
\cite{bauer,yasuda,scheidt,takeuchi2,takeuchi}. Thus, both
properties appear to be intrinsic. Are the low-temperature
power laws indicating line nodes affected by sample
differences? Which of the puzzling results are intrinsic? These
questions need to be addressed in order to understand the
physics of CePt$_3$Si and, in consequence, to have a more
complete picture of superconductors without inversion symmetry.

In order to shed light on all these issues we studied the
temperature dependence of the magnetic penetration depth of
CePt$_3$Si single crystals in terms of structural defects,
impurities and off-stoichiometry. The samples were grown by
different groups using different techniques, what allows us to
get deep into the sample-dependent problem. The results are
compared to those previously obtained in a high-quality
polycrystalline sample \cite{mine7}. The present study suggests
that the intrinsic properties of CePt$_3$Si are more affected
by structural defects than by slightly off-stoichiometry or the
presence of secondary phases. At low temperature, however, a
linear dependence of the penetration depth is observed for all
single crystals.

\section{Experiment details}

We measured four single crystals, labelled A-1, A-2, B-1, and
B-2. Crystals A-1 and A-2 were grown from polycrystalline
CePt$_3$Si samples using a mirror furnace. Crystals B-1 and B-2
were prepared by the Bridgman method, in which the growth is
extremely slow and the defect density is very low. It is worth
mentioning here that conventional annealing, that uses an
evacuated quartz tube, of Bridgman-grown crystals is not very
effective and sometimes even harmful. The superconducting
critical temperature of the crystals, given in
table~\ref{sampleprop}, is defined as the onset of the
diamagnetic behaviour in the penetration depth data.  All
samples had plate-like shapes with dimensions around
$0.5\times0.5\times0.4$ mm$^3$.

\begin{table}[b]
\caption{\label{sampleprop} Characteristics of the samples used
in this work. Single crystals grown by the mirror furnace
method are labelled as A-n and those grown by the Bridgman
technique as B-n.} \hspace{0.25cm} \footnotesize\rm
\begin{tabular*}{\textwidth}{@{}l*{15}{@{\extracolsep{0pt plus12pt}}l}}
\br
      Sample &  $T_c$ (K)  &     Host phase           &    Second phase     \\
\mr
\verb    A-1  &  0.75    &   Ce$_{1.14}$Pt$_3$Si$_{0.58}$     &  yes (rich in Si)  \\
\verb    A-2  &  0.73    &   Ce$_{1.61}$Pt$_3$Si$_{0.91}$     &  no  \\
\verb    B-1  &  0.79    &   Ce$_{0.99}$Pt$_3$Si$_{1.14}$     &  yes (rich in Ce)\\
\verb    B-2  &  0.79    &   Ce$_{1.04}$Pt$_3$Si$_{1.07}$     &  no  \\
\br
\end{tabular*}
\end{table}

 Backscattered electron images and both energy-dispersive (EDX)
and wavelength-dispersive (WDX) X-ray spectra of the crystals
were obtained with an Electron Probe Microanalyzer JEOL
Superprobe 8900R. EDX spectra were used for qualitative
analyses. The quantitative analyses of the WDX spectra were
performed with the package CITZAF, using highly pure Ce, Pt,
and Si standards and the Bastin's phi-rho-z correction method
\cite{armstrong}. The stoichiometric formulas of the crystals
are shown in table~\ref{sampleprop}. Penetration depth
measurements were performed utilizing a 12 MHz tunnel diode
oscillator in a dilution refrigerator running down to the
lowest temperature of 40 mK (see details in~\cite{mine7}).

\section{Results}

%
%
 \begin{figure}[b]
 \begin{center}
 \scalebox{1.}{\includegraphics{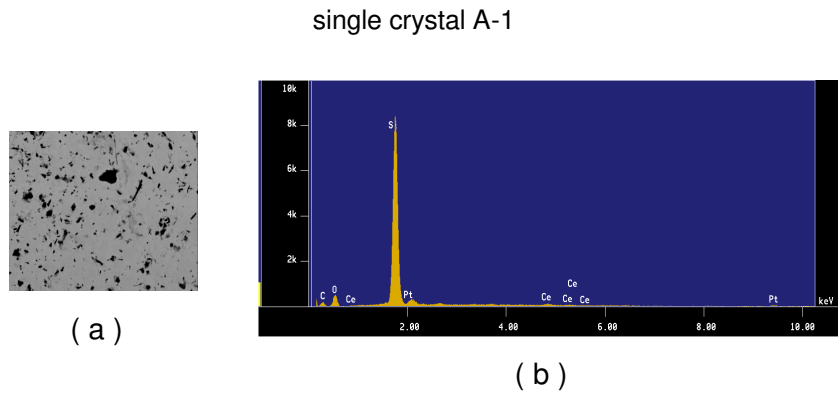}}
 \caption{(a) Backscattered electron image of a typical area of single crystal
 A-1. Horizontal side: 80~$\mu$m. The dark areas correspond to impurity phases.
 (b) Representative EDX spectrum
of the dark areas rich in Si.}
 \label{image1}
 \end{center}
 \end{figure}
%
%

%
%
 \begin{figure}
 \begin{center}
 \scalebox{1.}{\includegraphics{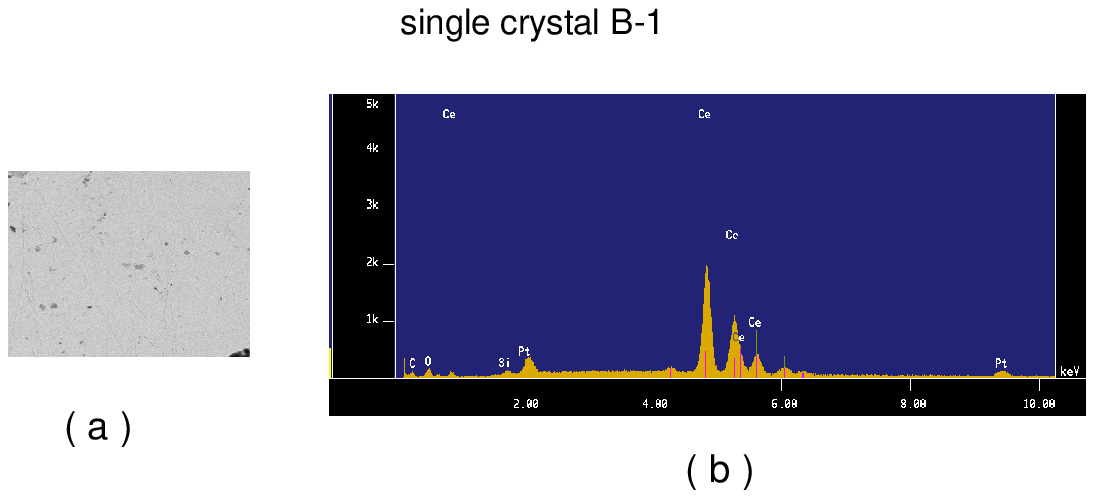}}
 \caption{(a) Backscattered electron image of a typical area of single crystal
 B-1. Horizontal side: 240~$\mu$m. The dark areas correspond to impurity phases.
 (b) Representative EDX spectrum
of the dark areas rich in Ce.}
 \label{image2}
 \end{center}
 \end{figure}
%
%

%
%
 \begin{figure}
 \begin{center}
 \scalebox{1.}{\includegraphics{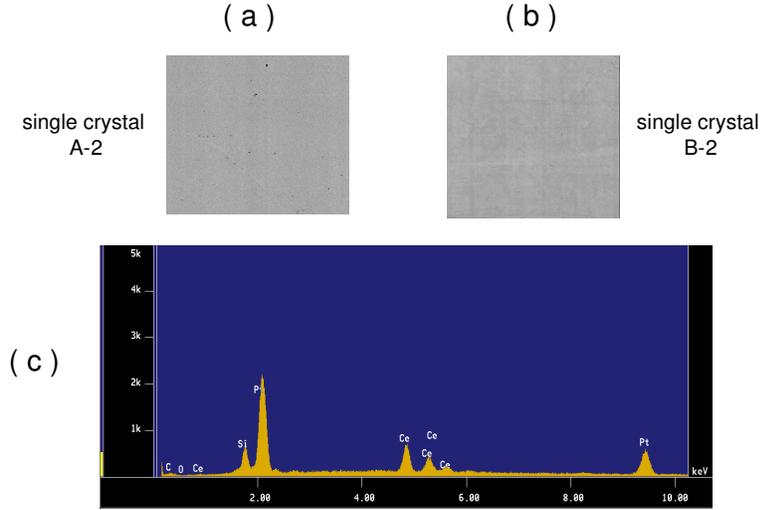}}
 \caption{Backscattered electron images of typical areas in single crystals
 (a) A-2 and (b) B-2. Horizontal side: 240~$\mu$m. No dark areas are observed
 in these images, implying single-phase crystals. (c) EDX spectrum of single
 crystal B-2.}
 \label{image3}
 \end{center}
 \end{figure}
%
%

Single crystals A-1 and B-1 had secondary impurity phases which
are seen as dark spots in the backscattered electron images of
typical sectional areas display in figures~\ref{image1}(a) and
\ref{image2}(a). The EDX spectra of the most abundant impurity
phases (dark zones) are shown in figures~\ref{image1}(b) and
\ref{image2}(b). The impurity phases of A-1 (B-1) were rich in
Si (Ce), and the host phases of both crystals were
off-stoichiometry (see third column of table~\ref{sampleprop}).
Single crystals A-2 and B-2 had pure phases, as can be deduced
from the very homogeneous backscattered electron images
depicted in figures~\ref{image3}(a) and \ref{image3}(b). Single
crystal B-2 had an almost nominal stoichiometry 1:3:1. The EDX
spectrum of crystal B-2 is shown in figure~\ref{image3}(c). It
is interesting to note that second impurity phases appeared in
samples grown by both the mirror furnace and the Bridgman
method.

\begin{figure}
 \begin{center}
 \scalebox{0.7}{\includegraphics{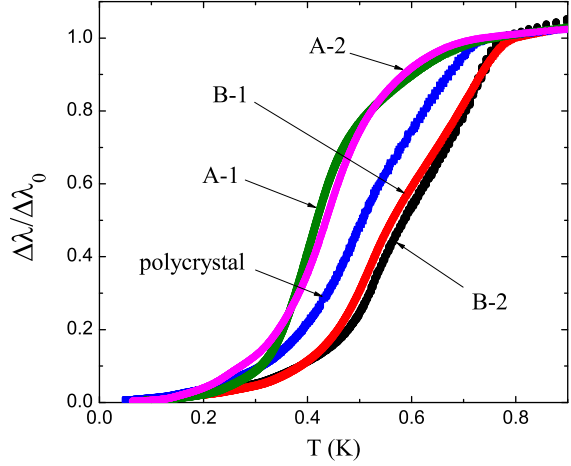}}
 \caption{Normalized in-plane penetration depth
 $\Delta\lambda_\parallel(T)/\Delta\lambda_0$ vs $T$ of our four single crystals of
CePt$_3$Si. For comparison, data of a polycrystalline sample (from~\cite{mine7}) are also shown.}
 \label{lambdatotal}
 \end{center}
 \end{figure}

The effect of growth process, defects, presence of impurity
phases, and off-stoichiometry on the superconducting properties
of CePt$_3$Si is analysed by means of the temperature-dependent
magnetic penetration depth. Figure~\ref{lambdatotal} shows the
normalized in-plane penetration depth
$\Delta\lambda_\parallel(T)/\Delta\lambda_0$ versus $T$ of all
four single crystals of CePt$_3$Si and, for comparison,
$\Delta\lambda(T)/\Delta\lambda_0$ of the polycrystalline
sample reported in \cite{mine7}. Here $\Delta\lambda_0$ is the
total penetration depth shift. Two features are noticeable in
this figure: (1) all onset temperatures are around 0.75 K (see
table~\ref{sampleprop}) and (2) the superconducting transitions
are broad.
%
%
%
%
The lower transition temperatures and much broader transitions
in samples A-1 and A-2, grown by the mirror furnace technique,
are most likely due to the presence of structural defects
and/or random impurities since crystal A-2 had no impurity
phases. Crystals B-1 and B-2 present higher critical
temperatures and less broad transitions than the high-quality
annealed polycrystalline sample studied in \cite{mine7} (see
table~\ref{sampleprop}). The penetration depth data of crystals
B-1 and B-2 are similar, even though crystal B-1 had second
impurity phases.

\begin{figure}
 \begin{center}
 \scalebox{0.7}{\includegraphics{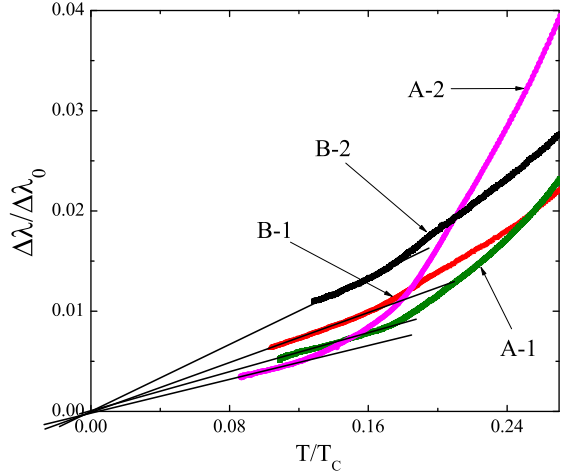}}
 \caption{Low-temperature region of normalized in-plane penetration depth
$\Delta\lambda_\parallel(T)/\Delta\lambda_0$ vs $T/T_c$ of our
four single crystals of CePt$_3$Si. A linear low-temperature
behaviour is observed in all crystals, as it was found
previously in a polycrystalline sample \cite{mine7}.}
 \label{ltlambda}
 \end{center}
 \end{figure}

To study the effect of sample quality on the line nodes of
CePt$_3$Si \cite{mine7,izawa}, we depict in
figure~\ref{ltlambda} the low-temperature region of
$\Delta\lambda_\parallel(T)/\Delta\lambda_0$ for our four
single crystals. The linear temperature dependence of the
penetration depth -indicating line nodes in the energy gap-
found in an annealed polycrystalline sample \cite{mine7} is
also detected in single crystals. The outstanding observation
in figure~\ref{ltlambda}, however, is that sample quality does
not affect such linear temperature behaviour. The line nodes of
CePt$_3$Si are robust regarding disorder and the mechanism
causing the broadening of the superconducting transition.
%
%
%
%

\section{Discussion}

The fact that the superconducting transition of CePt$_3$Si as
measured by penetration depth is wide, independently of growth
procedure, absence of impurity phases, and stoichiometry,
agrees with all previous results obtained by other techniques
in all kind of samples, varying from
unannealed/quenched/annealed polycrystalline samples to
annealed single crystals. The sharpest transitions, with a
width around 0.15 K, have been seen in annealed polycrystalline
samples \cite{motoyama2} and annealed single crystals grown by
the Bridgman technique \cite{takeuchi2}. For the latter ones it
was even possible to observe de Haas-van Alphen oscillations, a
clear indication of the low-disorder character of these samples
with a mean free path $l=1200-2700$~{\AA} much larger than the
coherence length $\xi(0) \sim 100$~{\AA}. We remark that a
width of 0.15 K is quite large considering the transition
temperature of 0.75 K. From our results we believe that
impurity phases and off-stoichiometry do not play significant
roles in the broadness of the transition.

\begin{figure}
 \begin{center}
 \scalebox{0.7}{\includegraphics{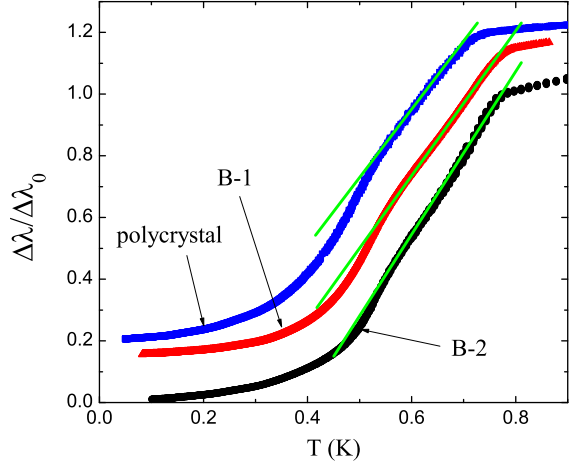}}
 \caption{Normalized in-plane penetration depth data of crystals B-1 and B-2,
 which have the sharper superconducting
transitions. Data of a polycrystalline sample (from~\cite{mine7}) are also shown.
Data of the polycrystalline and B-1 samples have been shifted for clarity.}
 \label{lambdacrystals}
 \end{center}
\end{figure}

Is the transition width related to the occurrence of an
inflection point around 0.5 K in penetration depth and a second
peak in some specific heat data? The answer seems to be no. In
specific heat measurements the transition is always broad,
independently of whether it occurs at 0.7 K
\cite{bauer,scheidt,jskim,motoyama2}, in which case sometimes
two peaks appear, or at 0.5 K
\cite{tateiwa,takeuchi2,motoyama2}, in which case only one peak
is always observed. To verify what is seen in penetration depth
measurements we plot again in figure~\ref{lambdacrystals} the
penetration depth data with the sharper superconducting
transitions, where the data of the polycrystalline and B-1
samples have been shifted for clarity. In our best sample (B-2)
the inflection point is barely observed, yet the transition is
wide. We notice that the second less stronger drop in the
penetration depth shows up independently of the existence of
second impurity phases. It is possible that in a slightly
better sample, with less defects, the second drop withers
whereas the transition remains broad. We argue here that the
second anomaly in either specific heat or penetration depth of
CePt$_3$Si does not have an intrinsic superconducting origin. A
second superconducting transition, suggested in previous
reports \cite{scheidt,nakatsuji}, is discarded. The transition
broadness could be an effect of lack of inversion symmetry;
however, it has not been observed in other noncentrosymmetric
superconductors.

%
%
%
%

Our penetration depth data display the superconducting
transition around 0.75 K. Is this the true superconducting
transition temperature of CePt$_3$Si or is it 0.5 K? We argue
that sample quality does not make a difference in this issue.
For example, for the same single crystal $T_c$ was 0.5 K when
probed by thermodynamic properties like specific heat and
thermal conductivity, but was 0.75 K when measured by
resistivity \cite{izawa,takeuchi2}. In general, $T_c$ of
CePt$_3$Si depends on how it is measured; that is, depends on
the measuring technique. The transition takes place around 0.5
K in thermodynamic measurements like specific heat
\cite{tateiwa,takeuchi2,motoyama2} and thermal conductivity
\cite{izawa} and in magnetic probes like NMR $1/TT_1$
\cite{yogi2}. It occurs at 0.75 K in inductive methods like
penetration depth and ac susceptibility
\cite{young,nakatsuji,takeuchi,motoyama}, and in resistivity
\cite{izawa,takeuchi2}. An odd result comes from a recent ac
magnetic susceptibility measurement that found a broad
transition at 0.5 K \cite{mota}. Interestingly, within each set
of measuring techniques given different $T_c$'s, the probes
have no physical connection with one another. This makes the
case of CePt$_3$Si different from that of the heavy-fermion
superconductor CeIrIn$_5$, in which resistivity measurements
show the superconducting transition around 1.2 K whereas
thermodynamic and magnetic properties display it at 0.4 K
\cite{petrovic1}. At present the origin of this inconsistency
in CePt$_3$Si is not clear to us.

To discuss the low-temperature response of the penetration
depth,  we consider that the absence of inversion symmetry in a
crystal structure causes the appearance of an antisymmetric
potential gradient $\nabla V$ that leads to an antisymmetric
spin-orbit coupling (ASOC) $\alpha (\textbf{k}\times\nabla
V)\cdot \hat{\sigma} $. Here $\textbf{k}$ is the electron
momentum and $\alpha$ a coupling constant. In superconductors a
strong ASOC lifts the spin degeneracy existing in
parity-conserving systems by originating two bands with
different spin structures and with energy gaps \cite{frigeri}
%
\begin{equation}
\label{split} \Delta_{\pm}(\textbf{k})=\psi \pm
t|\textrm{\textbf{g}} (\textbf{k})|
\end{equation}
%
Here $\psi$ and $t|\textrm{\textbf{g}}(\textbf{k})|$ are the
spin-singlet and spin-triplet components, respectively, and
$\textrm{\textbf{g(k)}}$ is a dimensionless vector $\left[
\textrm{\textbf{g(-k)}}=-\textrm{\textbf{g(k)}} \right]$
parallel to the vector $\textbf{d(k)}$ of the spin-triplet
order parameter. In this spin-split band model the order
parameter is then a mixture of spin-singlet and spin-triplet
states, and when the spin-orbit band splitting $E_{so}$ is much
larger than the superconducting energy scale $k_BT_c$ the
system can be essentially regarded a two-gap superconductor
with gaps that open at $T_c$. In CePt$_3$Si it is assumed
\cite{frigeri} that the gaps in equation~(\ref{split}) have an
isotropic $s$-wave spin-singlet component and a $p$-wave
spin-triplet component with $\textbf{g(k)} \propto (k_y,-
k_x,0)$. This spin-triplet component has point nodes.

\begin{figure}
 \begin{center}
 \scalebox{0.5}{\includegraphics{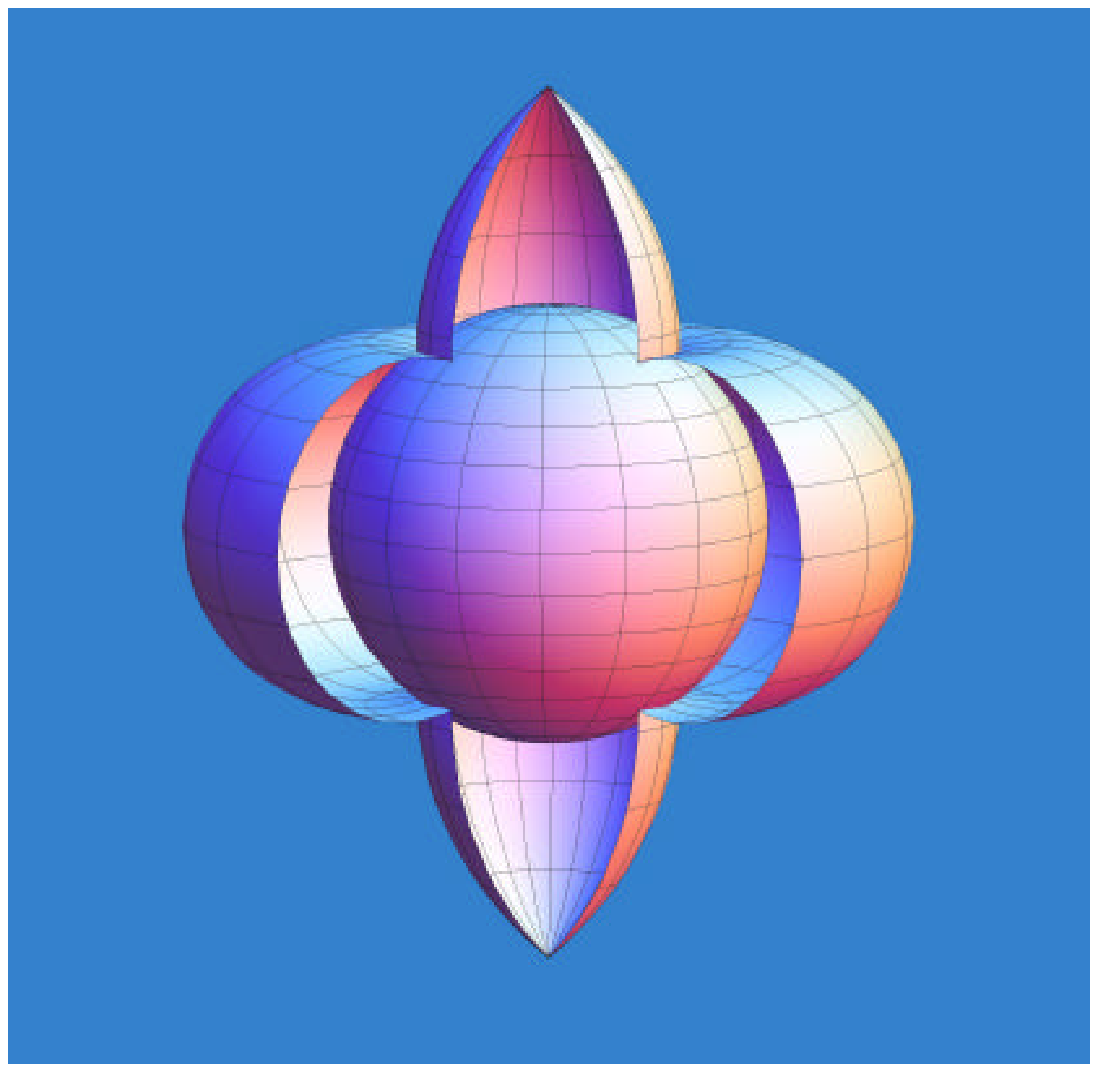}}
 \caption{3D polar plot of the gap function $\Delta_-(\textbf{k})$ in
CePt$_3$Si.}
 \label{gapminus}
 \end{center}
 \end{figure}

Line or point nodes in the energy gap structure are natural and
required by symmetry in some unconventional centrosymmetric
superconductors, but their existence is not obvious in
parity-violating superconductors. The line nodes in CePt$_3$Si
can be, however, confidently explained by using the spin-split
band model \cite{hayashi2}. They turn out to be accidental -not
imposed by symmetry- and appear in the gap
$\Delta_{-}(\textbf{k})=\psi - t|\textrm{\textbf{g}}
(\textbf{k})|$ when $(\psi/t) < 1$. For CePt$_3$Si, in
particular,  $(\psi/t) \sim 0.6$ and the energy gap
$\Delta_{-}(\textbf{k})$ has the 3D polar representation shown
in figure~\ref{gapminus}.

%
%
%
%
The broad penetration depth response below $T_c$ leads to a
suppression of the superfluid density in the high-temperature
region. To expound such a suppression and the line nodes the
spin-split band model would require that the band with gap
$\Delta_{-}(\textbf{k})$ has a normalized density of states
about 0.9 \cite{hayashi2}, whereas energy band calculations
indicate that the difference in the density of states of the
spin-split bands is of the order of 0.3 \cite{samokhin}. Thus,
at present the broadness of the superconducting transition
cannot be accounted for in this model (see \cite{hayashi2} for
details).

We now discuss the effect of impurities and other defects that
may exist in our crystals on the superconducting behaviour of
noncentrosymmetric CePt$_3$Si. As we mentioned above, the
transition broadness and lower critical temperatures in samples
A-1 and A-2 are probably related to defects and/or random
impurities. We note that low defect and/or nonmagnetic impurity
concentrations affect the low-temperature properties of
unconventional one-band superconductors with inversion
symmetry. In the case of symmetry-required line nodes of a
one-component order parameter, it was proposed that in the
unitary scattering limit there is a crossover temperature
$T^{\ast} \sim (\Gamma \Delta_0)^{1/2}$ from a high-temperature
linear to a low-temperature quadratic behaviour of the
penetration depth \cite{HiGo}. $\Gamma$ is a scattering rate
that depends on the impurity concentration. For strong
scattering, a relatively small concentration of defects would
cause a high value of $T^{\ast}$ without significantly
depressing $T_c$. This is indeed observed in, for example,
unconventional $d$-wave high-$T_c$ superconductors
\cite{bonn94}. In noncentrosymmetric CePt$_3$Si with line nodes
such a behaviour is not seen. If one assumes that the optimum
onset temperature in CePt$_3$Si is that of crystals B's
($T_{c0}=0.79$ K) and that the lower onset temperature of
crystal A-2 ($T_c=0.73$ K) is due to the present of
defects/impurities, one has for the latter sample a relative
$T_c$ suppression $(T_{c0}-T_c)/T_{c0}$ of about 8$\%$ and a
$T^{\ast} \sim 0.38T_c$ (using $\Delta_0/T_c = 2.15$ for a
line-node gap). For crystal A-1 the relative $T_c$ suppression
is 5$\%$ and $T^{\ast} \sim 0.31T_c$. According to these
estimations, in crystals A-1 and A-2 the penetration depth
should follow a $T^2$ behaviour below about 0.3$T_c$, which is
not observed in the data shown in figure~\ref{ltlambda}. The
penetration depth instead becomes linear below $T \approx
0.16T_c$ in A-1 and $T \approx 0.12T_c$ in A-2.

From the above analysis it seems that in parity-violating
CePt$_3$Si with line nodes defects or nonmagnetic impurities do
not perturb the low-temperature penetration depth in the same
manner as in unconventional centrosymmetric superconductors. We
notice here the observation of weak pair-breaking effects of
nonmagnetic-ion substitutions on Pt and Si sites
\cite{bauer2,kaldarar}, where rare-ion concentrations as high
as 6$\%$ and 10$\%$ do not destroy superconductivity. This
apparent insensitivity to disorder has resemblances to a
conventional superconducting phase and suggests that the
superconducting order parameter of CePt$_3$Si is not an even
basis function that transforms according to the nontrivial 1D
irreducible representations of $C_{4v}$ ($A_2, B_1,$ and
$B_2$). Recent theoretical works \cite{frigeri2,mineev2}
considered impurity effects on the critical temperature and the
density of states \cite{liu} of superconductors without
inversion symmetry. It was found that for some particular cases
impurity scattering leads to a functional form of $T_c$ that,
up to a prefactor, is the same as the one for unconventional
superconductor with inversion symmetry: $\ln(T_c/T_{c0}) =
\alpha\left[\Psi(\frac{1}{2})-\Psi\left(\frac{1}{2}-\frac{\Gamma}{2\pi
T_c}\right)\right]$. For CePt$_3$Si $\alpha \sim 0.9$,
therefore there is not much change in the estimations of
$T^{\ast}$ carried out in the paragraph above. In the analysis
of the local density of states it was concluded that a single
nonmagnetic impurity-induced resonant state near the Fermi
energy is likely possible in noncentrosymmetric superconductors
\cite{liu}. This implies that defects or nonmagnetic impurities
would perturb the low-T response of thermodynamic variables and
the penetration depth, a disturbance that is not observed in
our low-temperature $\lambda(T)$ data.

\section{Conclusions}

We reported on magnetic penetration depth measurements on
several single crystals of CePt$_3$Si grown by different
techniques. We discussed the effects of the growth processes,
presence of impurity phases, off-stoichiometry, and defects on
the superconducting phase of this compound. We found that:

(a) The presence of impurity phases and off-stoichiometry have
relatively low influence on the broadness of the
superconducting transition. On the other hand, the growth
process, related to the disorder in the samples, has more
impact in the transition widening. The superconducting
transition in CePt$_3$Si may be intrinsically wide.

(b) The anomaly around 0.5 K, observed in several
superconducting properties, appears to fade out in the
penetration depth data as the transition gets sharper.

(c) In penetration depth measurements the superconducting
transition occurs around 0.75 K, as also found in resistivity
and ac  magnetic susceptibility measurements. This critical
temperature contrasts the one at 0.5 K detected in specific
heat, thermal conductivity, and NMR data.

(d) Defects and/or impurities do not change the linear
low-temperature dependence of the penetration depth of
noncentrosymmetric CePt$_3$Si with line nodes, as they do in
other unconventional superconductors with line nodes in the
gap.

\ack We are grateful to M Sigrist and D Agterberg for
discussions. This work was supported by the Venezuelan FONACIT
(Grant number S1-2001000693) and the Austrian FWF (Grant number
P18054).


\section*{References}


\providecommand{\newblock}{}

\end{document}